\documentstyle[11pt,renewpasp,twoside]{article}
\input{epsf}
\begin{document}
\title{Why Do Disks Form Jets?}
\author{D. Lynden-Bell}
\affil{Institute of Astronomy, The Observatories, Madingley Road,
Cambridge, CB3 0HA, UK \& Clare College, Cambridge, UK} 
\begin{abstract}
It is argued that jet modelers have given insufficient study to the
natural magneto-static configurations of field wound up in the presence
of a confining general pressure.  Such fields form towers whose height
grows with each twist at a velocity comparable to the circular
velocity of the accretion disk that turns them.  A discussion of the
generation of such towers is preceded by a brief history of the idea
that quasars, active galaxies, and galactic nuclei contain giant black
holes with accretion disks.
\end{abstract}
\section{Introduction and History}
Before a child learns to run, it must learn to walk.  Before it learns
to walk, it must learn to stand.  I believe that those of us studying
jets and winds from rotating objects have been guilty of attempting to
run before we have understood how to stand.

Here I give an account of the natural sequence of static magnetic
structures that are obtained by the continued twisting of a local
poloidal magnetic field by an accretion disk.  My contention is that
had we understood such structures long ago, then we would have
considered the jets seen in nature to be an obvious consequence of the
twisting of the field by an accretion disk in the presence of an
external pressure.  If we imagine the accretion disk as turning all
the time, then we have to consider a sequence of equilibrium models in
which the total twist angles increase linearly with time.  This
sequence of models gives towers whose height increases with time at a
velocity of the order of the maximum rotation velocity in the disk.

Although I gave the first account of these growing towers in
Lynden-Bell (1996), the magnetohydrodynamic wind modelers were
already much more sophisticated and it is only recently that Li et al.
(2001) decided that such an elementary model may give a fundamental
clue to what is really going on.  Their numerical method was not
capable of exploring such models beyond the first two twists.  While
these are highly indicative of what happens, the many-twist picture is
best understood from the somewhat more realistic version of my 1996
paper given here.  The essentials of the process can be understood
without any detailed mathematics and with only the minimal knowledge
of magnetohydrodynamics contained in Alfv\'{e}n's theorem that
magnetic flux moves with the fluid.  Those primarily interested in the
magnetic towers model of jets may be advised to move direct to \S
2 unless they have an interest in the history of the subject.

I was originally invited to give an account of black holes in galactic
nuclei, a subject I had some hand in starting, but by now there are
those far better qualified to discuss the recent exciting data and
David Merritt has given a magnificent critical account here.  My
counter-suggestion of giving some account of the history of black
holes in galactic nuclei followed by my recent work  on jets was
accepted, so I shall now turn to history.

Since the Isaac Newton Group is running this conference there is no
better starting point than Newton's first Query in his {\it Opticks} of 1704:
\begin{quote}{\small And do not Bodies act upon Light at a distance and,
by their action, bend its Rays, and is not this action ({\it caeteris
paribus\/}) strongest at the
least distance?}
\end{quote}
In 1784, moving on by 80 years we get to the Reverend John Michell's
great paper that not only predicted giant black holes but told us how
massive they would be and how they would be found.  Michell was a
remarkable scientist, one of the founders of the geology of the
UK through his early stratigraphy.  He was the first to
suggest that earthquakes traveled as waves.  Other achievements of
his were the experimental demonstration of an inverse square law
between magnetic poles, the determination of the first luminosity
function (using the Pleiades), the statistical demonstration that
binary stars were gravitationally associated, a method for measuring
the motion of the Sun through nearby stars and the invention of the
apparatus for measuring $G$, the gravitational constant.  This was
perfected and first successfully used after his death by his great
friend Henry Cavendish.   Erasmus Darwin, grandfather
of Charles, was among Michell's more prominent students at Cambridge.
Michell in a paper in the Philosophical Transactions of the Royal
Society for 1784 argued that
bodies of greater than 500 times the diameter of the Sun and not
of lesser density would so attract the light that it could not escape,
rendering such great masses invisible to our senses.  Nevertheless,
he thought, they might be detected through observations of small satellites
circling about them.
It is interesting that $(500)^3\ M_\odot = 1.25 \times 10^8\ M_\odot$ (Mitchell
1784),
while the first definitive observation of a giant black hole by Miyoshi
et al. (1995) used material circling around it to determine its mass---$3.6
\times 10^7\ M_\odot$.  Such predictions more than 200 years
in advance of their time are extremely rare!

We know that Michell's idea traveled to the French academy not only
by the customary exchange of journals but also it was discussed in an
exchange of letters between the botanist Sir Joseph Banks, President of
the Royal Society, and Benjamin Franklin, who was then (1783) US Ambassador
in Paris. Two years after Michell died, Laplace
described the idea of dark stars (\mbox{without} attribution) in his
{\it Syst\`{e}me du Mond}\footnote{Available in English translation
(Laplace 1809).}  Laplace (1795), using exactly Michell's argument translated into
French, but in 1802 Young published his two-slit experiment
demonstrating the wave nature of light (Young 1802).  
This put in doubt Michell's
arguments, based on the corpuscular theory of light, so Laplace
deleted the passage on dark stars from the second edition of his book.

The well known historian of science Agnes Clarke, writing an
account of Michell's life and work for {\it The Dictionary of National
Biography} (Clarke 1917),
ended her list of Michell's scientific achievements with the words 
\begin{quote}
{\small But [he] speculated fruitlessly on a supposed retardation of
light through the attraction of its corpuscles by the emitting masses.} 
\end{quote}
This almost certainly reflects the received wisdom of the British
astronomers of the time.

Returning to the 1780s, Sir William Herschel (1786) remarked upon the
bright nuclei of the galaxies we now know as NGC 4151 and NGC 1068 and
in 1850 Lord Rosse, using the giant telescope he had developed at Birr
Castle, discovered spiral structure in M51 and other galaxies (Rosse 1850).

Coming to the 20th century, Vesto Slipher (1917), taking long time-exposed
spectra photographically with a 36 inch telescope, discovered broad
emission lines in these galaxies.  They were so broad that he decided
some non-Doppler broadening mechanism must be involved.  It is seldom
pointed out that most of the early redshifts of galaxies are due to
Slipher's work.  He did not claim that the Universe was expanding
because he thought similar work in the southern hemisphere might find
almost nothing but blueshifts.  However, his work was used by
Eddington (1922) in his book on general relativity.

In 1918 Curtis photographed the jet at the centre of M87 (Curtis 1918).  
Even the
emission mechanism remained a mystery until Baade showed it to be
polarized synchrotron emission in 1952.  Synchrotron emission was
first worked out by Schott (1912) but it was not recognized as an
emission mechanism in astronomy until the 1950s.

The major postwar development of astronomy came out of British
developments in RADAR.  In the years 1955--1975 Sir Martin Ryle led his
Cambridge team to a series of remarkable discoveries which impacted
almost all parts of the subject from stellar death to cosmology (Ryle
1968).  Thus the third Cambridge Catalogue of radio sources and their
3C numbers became deeply embedded in the subject.  Ryle and Hewish together
developed aperture synthesis as a tool.  In this they were helped not
a little by a secret weapon.  David Wheeler in Cambridge had developed
a program for inverting Fourier transforms which was many times faster
than others then in use.  When some ten years later the method was
rediscovered and published by Cooley and Tukey  it rightly became
famous.

Henry Palmer of Jodrell Bank led a team that developed linked radio
telescopes over considerable distances and showed that a small
fraction of the 3C sources contained small-angular-diameter sources
down to a few arcseconds.  Hazard realized that an accurate position
for one of these could be obtained from a lunar occultation that would
take place in Australia (Hazard, Mackay, \& Shimmins 1963).  
Such positions were vital for the
identification programs which Schmidt and Sandage had started at the
Mt Wilson and Palomar Observatories.  The first quasar spectrum to be
taken was that of 3C48 by Greenstein.  This showed emission lines at
wavelengths that defied identification.  Later that year (1963) Hazard
sent Schmidt the lunar occultation position of 3C273.  Luckily this
showed two lines whose wavelength ratio exactly matched those of the
hydrogen spectrum at the amazing redshift of $z=0.158$, quite unheard
of for a 12$^{{\rm th}}$ magnitude object.  

Schmidt's (1963) identification 
was soon confirmed by Oke's (1963) detection of a third hydrogen line
in the infrared.  With the clues provided by 3C273, Greenstein's
spectrum of 3C48 was then interpreted as having a redshift of $z =
0.367$.  Soon afterwards, Greenstein \& Schmidt (1964) collaborated
on a paper that demonstrated that the high redshifts of the quasars
could not be of gravitational origin.  The emission lines needed a
large volume for their formation and in a strong gravity field this
would involve much greater line broadening than the emission lines
showed.  Hoyle and Fowler tried to get out of this by putting the
source of emission at the centre of a very massive large stellar
system which provided the potential but such a contrived model
acquired few backers.

Early attempts to find cluster galaxies associated with quasars failed
and this let to the idea that quasars were not associated with
galaxies at all.  Novikov (1964) and Ne'eman (1965) independently put
forward the idea that they were white holes---time-delayed pieces of
the Big Bang.  It was Salpeter who first considered a massive object
ploughing through a galaxy and accreting interstellar gas.  Assuming
that the material gradually percolated down to the last stable
circular orbit, he gave the 5.7\% efficiency of turning mass into
energy.  The properties of the orbits in Schwarzschild's metric had
indeed been known to relativists for some years, Zel'dovich among them.

Three papers laid the foundation of the thin accretion disk model now
widely used

\begin{itemize}
\item Salpeter, E.E.  1964, `Accretion of interstellar matter by
massive objects', ApJ, 140, 796
\item Lynden-Bell, D.  1969, `Galactic nuclei as collapsed old quasars',
Nat, 223, 690
\item  Bardeen, J. 1970, `Kerr metric black holes', Nat, 226, 64

\end{itemize}

G. Burbidge (1959) gave minimum energy estimates of order $10^{61}$
erg for a number of radio sources and rightly emphasized that this 
was a very large number,  while well before quasars Ambartsumian had
emphasized the extraordinary behavior and high energy emissions from
galactic nuclei (Ambartsumian \& Shakbuzian 1958).

The main points of my 1969 paper were: 

\begin{enumerate}
\item $10^{61}$ erg weigh $\onehalf 10^7\ M_\odot$.  If these ergs
arise from nuclear fusion then $10^9\ M_\odot$ must be involved.  But
quasars vary in as little as 10 hours and $G(10^9\ M_\odot)^2/(10\ {\rm
light\ hour}) \gg 10^{61}\ {\rm erg}$.  The assumption that the
energy is nuclear therefore leads to an even greater gravitational
binding energy which must have been lost.  Thus the assumption is
wrong.  

Hence most of the power of quasars is gravitational in origin and
masses $\sim 10^8\ M_\odot$ suffice.  But there are no dead states like
white dwarfs for such masses, except black holes.  If emission is not
100\% efficient remnants must remain.

\item I next considered the number of dead QSOs---based on Sandage's
(1965) estimates of them---and obtained

\[
\left( \begin{array}{c}
{\rm number\ of\ clusters}\\ {\rm of\ galaxies}\end{array}\right) <
\left( \begin{array}{c}
{\rm number\ of}\\ {\rm dead\ quasars}\end{array}\right) <
\left( \begin{array}{c}
{\rm number\ of}\\ {\rm galaxies}\end{array}\right)
.\]
\item
How do we hide dead quasars of $\sim 10^8\ M_\odot$ when they still
gravitate?  They are likely to be in dense places and to surround
themselves with stars.  Thus the obvious place to find them is in
galactic nuclei.

\item
What should dead quasars look like?  Taking them to accrete via a
friction caused by magnetic torques, I derived a temperature
distribution $T(r) \propto r^{-3/4}$ and adding many such rings of
black hole body emission gave me
$$S_\nu \propto F_M^{2/3} \nu^{1/3} \exp - (h\nu/kT_{\rm max}),$$
where $F_M$ is the rest-mass flux down the hole.

\item
Observational predictions were that most large galaxies should have
nuclei with high M/L when inactive.  In calling the paper `Galactic
nuclei as collapsed old quasars', I was raising the question ``Are
galactic nuclei just stars gathered around such black hole remnants of
quasars?'' 
\end{enumerate}

The masses of the black holes estimated in 1969, mainly from Merle
Walker's observations using the Lallmand image-tube, are compared with
modern values in Figure 1.  It is clear that these old estimates are
mostly too high by a factor of order ten because no allowance was made
for the mass of stars in the nuclei.

\begin{figure}[!ht]
\plotone{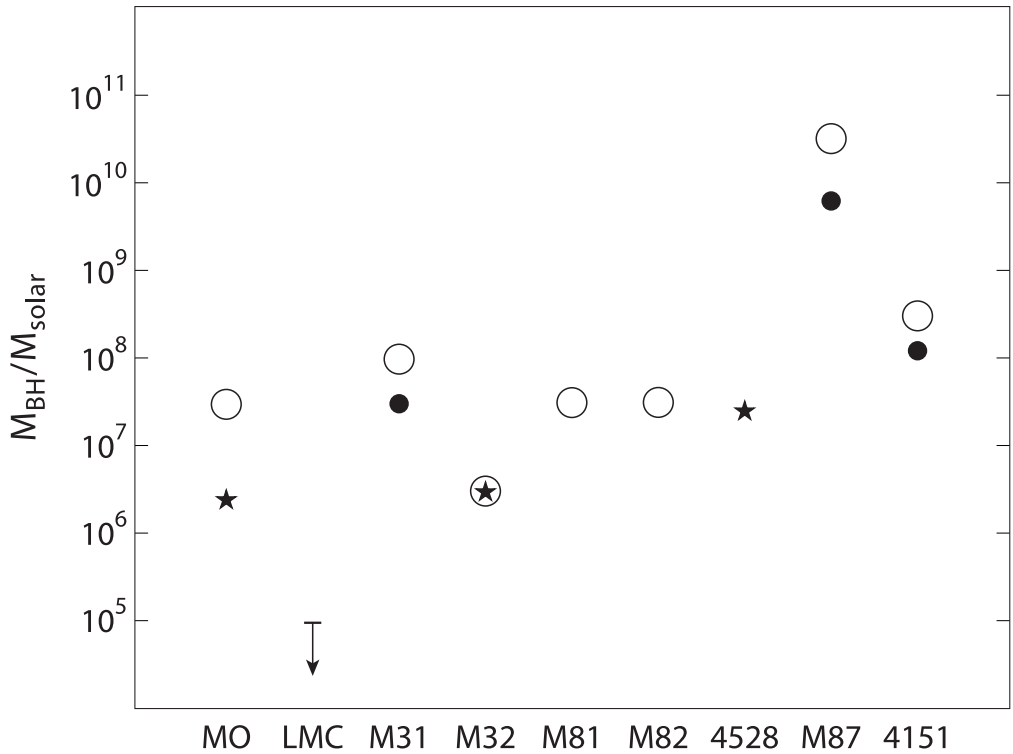}
\caption{My 1969
estimates of Black Hole masses $\circ$, compared with modern
determinations - $\star$ good, and $\bullet$ less definitive.}
\end{figure}

At the time, most working astronomers were highly sceptical of the
whole picture, but Maarten Schmidt volunteered that it seemed the best
theory available to explain the amazing phenomena.

It was clear that as a black hole accreted from the last stable orbit
it would accrete angular momentum as well as mass.  It thus became
important to work out accretion disks with Kerr holes.  I had a
friendly competition with Jim Bardeen to do this and his knowledge of
general relativity gave him a head start which he easily kept.  His
short but beautiful paper (Bardeen 1970) shows how a Schwarzschild
hole evolves into an extreme Kerr hole as it accretes.  Some of this
is covered in my Vatican review (Lynden-Bell 1971a) 
and radio observations of the
Galactic Centre were made then (Ekers \& Lynden-Bell 1971).

In 1971 I reviewed the data on the Galactic Centre with Martin Rees
(Lynden-Bell \& Rees 1971)
and some further details of Kerr disks and vortices were given later
(Lynden-Bell 1978, 1986), but by then the field had become overpopulated.  

A maximum likelihood method for determining luminosity functions was
developed and applied to quasars and miniquasars (Lynden-Bell 1971b,c).
\section{Jets from Accretion Disks} 
Jets were not generally acknowledged to feed radio lobes before Rees's
(1971) theory.  Jets are seen in nature associated with radio galaxies
(M87, Cygnus A; Hargrave \& Ryle 1974), quasars (3C273, 3C47),
young stars (Beckwith et al. 1990; Bouvier 1990)
Herbig--Haro objects (Bontemps et al. 1996; Burrows et al. 1996), dying
stars (SS 433), and micro-quasars (Mirabel \& Rodriguez 1999).

One common feature is that the jet velocities are of the order of the
circular velocity that would balance gravity at the central object's
surface.  

Theoretical models of jets fall into three classes:

\begin{enumerate}
\item
Hydrodynamic models collimated by Laval nozzles (Blandford \& Rees
1974), or by vortices around black holes (Lynden-Bell 1978), or by
self-similar thick disks (Gilham 1981; Narayan \& Yi 1995; Narayan, Barret,
\& McClintock 1997).

\item
Wind models in some way collimated by the local magnetic field of the
rotating source (Mestel 1968; Blandford \& Znajek 1977; Sakurai 1985, 
1987; Heyvaerts
\& Norman 1989; Appl \& Camenzind 1993; Lery et al. 1998;  Lynden-Bell
1996; Okamoto 1997, 1999; Li et al. 2001).

\item
Models collimated by a large-scale pre-existing magnetic field
(Lovelace 1976; Blandford \& Payne 1982; Shibata \& Uchida 1985, 1986;
Begelman \& Li 1994; Bell \& Lucek 1995; Lucek \& Bell 1996; Ouyed \&
Pudritz 1997; Ouyed, Pudritz, \& Stone 1997; Vlahakis \& Tsinganos 1998).

\end{enumerate}

Models in class 3 certainly work but their collimation is no surprise
since the large-scale magnetic field provides collimation at great
distances.  A body of plasma initially fired across the field at some
angle $\theta$ to it would have its transverse motion resisted by the
field (a high conducting background medium being assumed) but its
$\cos \theta$ motion along the field would continue.  Eventually it would
be collimated to move along the field.  Thus, to some degree, all such
models provide collimation by fiat in the boundary conditions.  They
cannot provide collimation in dying stars or bodies such as SS 433 in
which the jets precess (Margon 1984).

The prevalence of electromagnetic phenomena associated with all jets
and the acceleration of some particles to very high energies with
relativistic motions strengthen the idea that the jets themselves may
be an electromagnetic phenomenon and even the early advocates of
purely hydrodynamic jets have more recently turned to
magnetohydrodynamic models.  This is partly because of the difficulty
of getting sufficiently high speeds but also because it was difficult
to obtain the extremely narrow jets like those seen in Cygnus A and
HH 30.

Here we shall therefore concentrate on the models in class 2, in which
the jets are collimated not by external fields but by fields
associated with the objects themselves.  Work in this area began from
consideration of stellar winds and the angular momentum that they
transported thereby braking stellar rotation, but it took on a new
life in the discussion of pulsars.   Following Sakurai's early work
there has been much discussion of asymptotic collimation of such
winds, however the whole subject has been put in a well ordered form
in Mestel's (1999) book so the interested reader may find it there.  I gave a
brief review of the still active controversies in my introduction to
the Royal Society's meeting on {\it Magnetic Activity in Stars, Disks and
Quasars} (Lynden-Bell et al. 2000). 

As stated in the introduction, we shall now turn to my growing-towers
picture (Lynden-Bell 1996) based on motion through a sequence of
static structures that inevitably occur when a poloidal magnetic field
is progressively twisted.

\subsection{Magneto-Static Theorems and Rough Estimates}
We consider a magnetic field anchored on the plane $z=0$ and confined
by an ambient external pressure $p_0$ due to an ionized medium.  Where
there is field there is no gas pressure and except at the surface of
the volume $V$ occupied by field (and $z=0$) the field configuration
is force free.  The total energy of the configuration may be written
$$W=W_R+W_\phi +W_z + W_p = \frac{1}{8\pi} \int \left (B^2_R +
B^2_\phi + B^2_z \right ) dV +p_0V. \eqno (1)$$
We have taken cylindrical polar components of the field for our later
convenience but the theorems proved are not confined to fields of any
particular symmetry.  Now the field wiggles its way to the minimum
energy configuration subject to the constraints imposed by the
boundary conditions on the disk.  These define the vertical field
component $B_z$ on $z=0$ and the twist angles of each field line which
are most easily conceived for the axially symmetric case in which they
are measured as twists about the axis.  It may be shown that the force
free equations follow from minimizing the energy subject to the
constraints.  
\newtheorem{Theorem1}{\sc Theorem}
\begin{Theorem1}
For any finite magneto-static configuration anchored on {\rm z}  = 0 then on
any cut  {\rm z} = constant
$$w_R + w_\phi = w_z + w_p, \eqno (2)$$
where ${\rm w}_{{\rm R},\phi,{\rm z}}$ {\it =} $[{\it 1/}({{\it 8}\pi})]\int {\rm B}^2_{{\rm R},\phi, {\rm z}} {\rm dA}, \  {\rm w_p} =
{\rm p_0A}$ and {\rm A} is the area of the volume {\rm V} intersected by the cut.  

Notice that the capital {\rm W} are related to the small {\rm w} by
integration over {\rm z}.
\end{Theorem1}
\noindent
{\sc Proof\hspace{1mm}}
Consider the slice between the cut at $z$ and one at
$z+dz$.  Lift the whole field configuration above $z+dz$ rigidly by an
amount $(\lambda -1) dz$, expand the slice to thickness $\lambda dz$
and leave the configuration below $z$ unchanged (see Figure 2).  In the
vertical expansion of the slice flux conservation gives 
$$B_R \rightarrow B_R/\lambda\ , \ \ B_\phi \rightarrow
B_\phi/\lambda\ , \ \ B_z \rightarrow B_z \ , \ \ dAdz \rightarrow
\lambda d Adz.$$
Thus in terms of the old fields and old $w$
$$\Delta W = \left [(w_R + w_\phi)\ (\lambda^{-1}-1) + (w_z + w_p)\
(\lambda -1)\right ] dz. $$
Now the original situation must be a minimum energy one, so $\partial
(\Delta W)/\partial \lambda$ must be zero at $\lambda =1$.  This
condition yields at once $w_R + w_\phi = w_z + w_p$ (QED).
\newtheorem{Theorem2}[Theorem1]{\sc Theorem}
\begin{Theorem2}
For any finite magneto-static configuration anchored on z {\it = 0}
$$W_z =W_p +\onehalf W_0,\eqno(3)$$
where ${\rm W_{\it 0}}$ {\it =}  ${\rm [{\it 1/}({\it 4\pi})] \int B_z B_RRdA}$ evaluated on {\rm z} {\it = 0}.
\end{Theorem2}
\noindent

\begin{figure}
\plotone{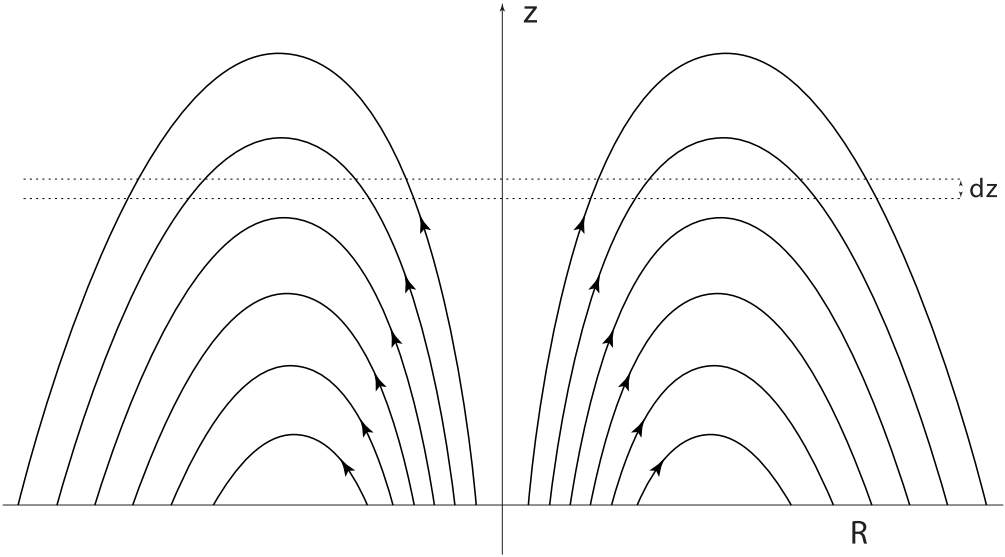}
\caption{A slice of a force-free field structure.  The twists about
the axis are not shown.  The theorem is true without assumptions of
axial symmetry.}
\end{figure}

\noindent {\sc Proof\hspace{1mm}} Consider a lateral expansion of the volume $V$ occupied by field such
that $R\rightarrow \mu R,\ z\rightarrow z$.  Conservation of flux
across elementary areas $Rd\phi dz$, etc., gives $B_R\rightarrow
B_R/\mu,\ B_\phi \rightarrow B_\phi/\mu,\ B_z \rightarrow B_z/\mu^2,\
dA\rightarrow \mu^2dA$
$$\Delta W=W_z(\mu^{-2}-1)+W_p(\mu^2-1).$$
Now we must take care because, unlike our former transformation, this
one moves the anchored foot-points on $z=0$.  Luckily the principle of
virtual work comes to our aid.  We have to add to $\Delta W$ the
work done in infinitesimal movement $\delta \mu R$ of the foot points
by the forces of constraint, viz. $B_RB_zdA/(4\pi)$, so 
$$\left [\frac{\partial \Delta W}{\partial
\mu}\right]_{\mu=1}+\frac{1}{4\pi}\int_{z=0} B_RB_zRdA = 0.$$
Thus we obtain
$$W_z = W_p +\onehalf W_0$$
(QED).
To those who believe that the pinch effect needs no external pressure
it comes as a surprise that $8\pi W_\phi = \int B^2_\phi \mu^{-2}\mu^2
dAdz$, which is independent of $\mu$.  Thus there is no tendency for {\sl
overall} contractions in a purely toroidal field.  It is perhaps of
interest to minimize the energy of a toroidal flux $F$ contained
between cylinders of radii $R_1$ and $R_2$ of height $Z$.  One finds
$B_\phi \propto {1}/{R}$:
$$W_\phi = \frac{1}{4} \frac{F^2_\phi Z^{-1}}{{\rm ln}(R_2/R_1)}.$$
Thus if $R_2$ and $R_1$ are increased by the same factor there is no
change in $W_\phi$, in agreement with what we found above, but if $R_2$
is fixed then $W_\phi$ decreases as $R_1$ is reduced giving rise to a
pinch.  In fact $B^2_\phi \propto {1}/{R^2}$ shows that the
toroidal magnetic field acts as a pressure amplifier delivering on
$R_1$ a pressure $(R_2/R_1)^2$ times that exerted on $R_2$.
The whole pinch effect is a pressure amplifier---without anything to
push on outside it has nothing to amplify and indeed the field would
expand outwards as well as inwards.

To orient ideas it is useful to make some very crude dimensional
estimates.  Suppose the typical radius out to which the field expands
laterally is $R_2$ and we deal with an axially symmetrical field
configuration which is wound up by the turning of the accretion disk.
Let the field pass upward through the disk at small radii and return
downward at larger radii and let the total upward poloidal flux be
$F$.  With each turn of the upward flux relative to the downward a
toroidal flux equal to $R$ will be generated, so after $n$ turns the
toroidal flux will be $nF$.  If $Z$ is the total height of the
configuration then the volume $V$ occupied by field will be $V=\pi
R^2_2Z$ a typical value of $B_\phi$ will be $nF/(R_2Z)$, a typical
value of $B_R$ will be $F/(\sqrt{2}\pi RZ)$, and a typical value of
$B_z$ will be $2F/(\pi R^2_2)$;  the $\sqrt{2}$ in $B_R$ comes from the
fact that the typical $\pi R$ for a radial field line is
$2\pi(R^2_2/2)^{1/2}$, while the factor 2 in the $B_z$ comes from the
fact that the flux goes both up and down.

Dividing (2) by $A/(8\pi)$ yields the interesting exact result
$${\Big\langle} B^2_R{\Big\rangle} + {\Big\langle} B^2_\phi{\Big\rangle} = 
{\Big\langle} B^2_z{\Big\rangle}  + 8\pi p_0, \eqno(4)$$
where the averages are taken over any $z={\rm constant}$ plane.
Putting in our rough estimates we get
$$\frac{F^2}{2\pi^2R^2_2Z^2}\left(1+2\pi^2n^2\right
)=\frac{4F^2}{\pi^2R^4_2} +8\pi p_0.\eqno(5)$$
 Now consider the case of no bounding pressure.  Then 
$$\frac{Z^2}{R^2_2}  = \frac{1}{8} \left (1+2\pi^2 n^2 \right), \eqno(6)$$
so ${Z}/{R} \rightarrow ({\pi}/{2})n$
 for $n$ large,
which appears to show that the collimation $Z/R$ increases with every
turn $n$.  On first finding this I was much encouraged and immediately
turned to get a more exact model.  When I solved it I was deeply
disappointed because before the field has twisted by even one turn it
relieves itself by expanding to infinity with an opening angle of
$120\deg$---very far from the sort of collimation we seek.  Figure 3
shows the field configurations for twist angles of $130\deg$ and
$180\deg$.  When $208\deg = 2\pi/\sqrt{3}^c$ is reached the field is
outwards along cones up to $60\deg$ to the equator.  This disconnection
at infinity is precisely what was predicted by some pretty theorems of
J.J. Aly (1995) but, although it provides a beautiful illustration of these, it
fails to give the accurate collimation we seek.  However, those
interested in galaxies with conical outflows should perhaps pay more
attention to this magneto-static result described in detail in
Lynden-Bell \& Boily (1994).  I was so disappointed in the result,
which was initially found in 1979, that I laid the whole subject aside
and only returned to it in 1994.  It took two more years before I
realised that victory might still be snatched from the jaws of
defeat.  Expression (6) failed to give collimation only because $n$
could not be made large without the field disconnecting at infinity.
If a way could be found to prevent the field reaching out to infinity
then perhaps $n$ could be made large after all.  This led me to
introduce the ambient pressure $p_0$ to confine the field.  It then
takes an infinite amount of work for the field to extend to infinity
so this will not  occur at finite twist angles.  Let us now turn to
equation (3) and insert our rough estimates.  On dividing by $V/8\pi$
we find 
$$\frac{4F^2}{\pi^2R^4_2} = 8\pi p_0 + 2\sqrt{2} \frac{F^2}{\pi
R^2_2Z^2}, \eqno(7)$$
where we have left the $B_z$ field on the disk at $B_{z0}$ since this
is fixed but used our estimate of $B_R$ there too.  Eliminating $p_0$
between (7) and (5) we find
$$\frac{Z^2}{R^2} = \frac{1}{16}\left (2\sqrt{2} +1 + 2\pi^2 n^2\right),$$
so asymptotically $$Z/R \rightarrow \frac{\pi}{2\sqrt{2}}n,$$
just a factor $\sqrt{2}$ less than our first wrong estimate without
$p_0$.  With $Z^2/R^2$ large the last term in (7) is negligible
compared with the first, so $W_0$ may be neglected compared with
$W_z$.  Furthermore, with $n$ large $B_R$ becomes negligible compared
with $B_\phi$.  Thus highly wound configurations become tall 
towers. For such we have from (4) at each height
$${\Big\langle} B^2_\phi{\Big\rangle}  = {\Big\langle}
 B^2_z{\Big\rangle} + 8\pi p_0. \eqno(8)$$
Furthermore, we may consider local transformations that expand
$R\rightarrow \mu(z)R$ where $\mu(z)$ is slowly varying but becomes one
except in a region around some chosen height.  Even after this
transformation the fields will still have $B_R$ small so its square is
negligible compared with $B^2_z$.  From such transformations, which
leave $W_0$ unchanged, we can deduce that at any great height (3)
may be replaced by $w_z = w_p$, so on dividing by $A/8\pi$
$${\Big\langle}B^2_z{\Big\rangle} \approx 8\pi p_0. \eqno(9)$$
Hence from (8) we have everywhere well above the base $z=0$
$$\onehalf{\Big\langle}B^2_\phi{\Big\rangle} = {\Big\langle}B^2_z{\Big\rangle} = 8\pi p_0,\eqno(10)$$
where the averages are taken over areas at constant $z$.
\begin{figure}[!t]
\plotone{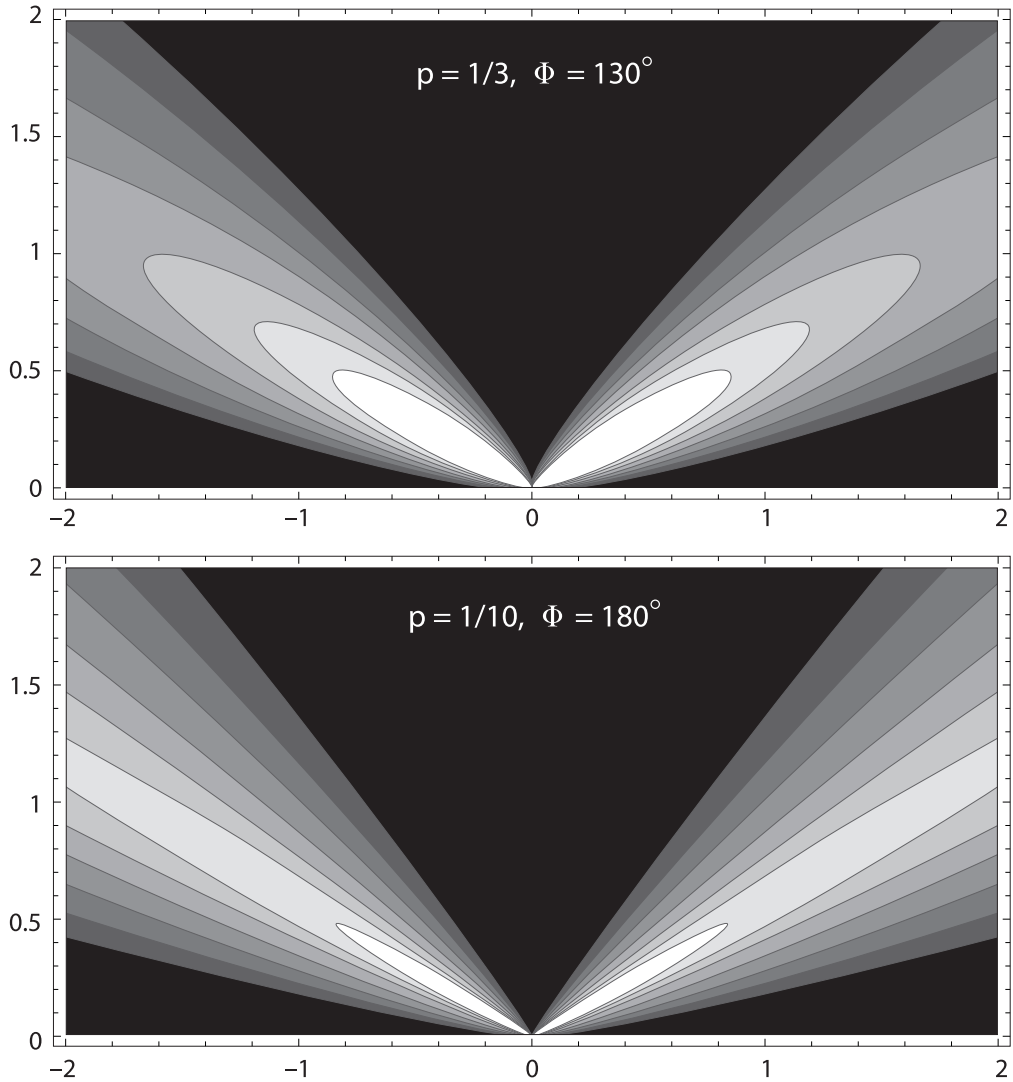}
\caption{Field lines splay out and reconnect at infinity after a twist
of only $2\pi/\sqrt{3}$ rad $\approx 208\deg$ unless confined by an
ambient pressure.} 
\end{figure}

\section{Better Estimates of the Field Structure}
Consider the tube of force that intersects the accretion disk in the
circle of radius $R_1$.  Let the poloidal flux rising within this tube
be $P$.  Then we may label each field line by the value of $P$ and indeed
$B_z = +[{1}/({2\pi R})] \partial P/\partial R$, $B_R = {-[1}/({2\pi
R})]{\partial P}/{\partial z}$.  If this field line returns to the
disk at some larger radius $R_2$ then there will be a differential
twisting due to the fact that on the accretion disk $\Omega_1 >
\Omega_2$.  We define $\Omega(P) = \Omega_1 - \Omega_2$ so $\Omega(P)$
is the rate of twisting of the field line labeled $P$.  Now suppose
that at any given time $t$ when the twist angle will have accumulated
to be $\Phi (P) = \Omega(P)t$ the field line labeled $P$ rises to a
maximum height $Z(P)$.  Then the total twist per unit height will be
$\Phi(P)/Z(P)$.  Now one turn (a twist of $2\pi$) will cause a
toroidal flux between $P$ and $P + dP$ of just $dP$ so the toroidal
flux per unit height between $P$ and $P + dP$ will be
$$\frac{\Phi(P)}{2\pi Z(P)}.\eqno(11)$$
Now only those field lines with $P^\prime \leq P$ will reach to height
$Z(P)$ so the total toroidal flux per unit height at $Z$ will be 
$$\int^{R(P)}_0 B_\phi dR = \int^P_0 
\frac{\Phi(P^\prime)}{2\pi Z(P^\prime)} dP^\prime$$
so
$$\bar{B}_\phi = \frac{1}{2\pi R(P)}\int^P_0
\frac{\Phi(P^\prime)}{Z(P^\prime)}dP^\prime, \eqno(12)$$
where $P$ is the flux label of the line that just reaches height
$Z(P)$ and no further.  The radius of the area occupied by magnetic
field at height $Z$ we call $R(Z(P))$, or $R(P)$ for short.  This is
not the radius at which the line labeled $P$ reaches height $Z$.

Now the flux $P$ is all the poloidal flux that crosses height $Z$ once
on the way up and once on the way down, so the average $\langle\vert
B_z\vert \rangle$ is $2P/ \left\{\pi \left[R(P)\right]^2\right\}$.  
Since $B_z = (2\pi R)^{-1}\partial P/\partial R$ we may write
$${\Big\langle} B^2_z{\Big\rangle} = \int^{R(P)}_0 
\left (2\pi R \right )^{-1}
        \left (\partial P/\partial R \right )^2dR = 
        \left (2P/
                \left\{\pi
                        \left [R(P)
                        \right]^2
                \right\}
        \right )^2 I,\eqno (13)$$
where $I$ is the dimensionless integral given by 
$$I=\frac{\pi}{8} \int^1_0 \left (\partial
\tilde{P}/\partial\tilde{R}\right )^2
\tilde{R}^{-1}d\tilde{R}\eqno(14)$$
and 
$$\tilde{P}=P^\prime(R^\prime)/P,\ \tilde{R} = R^\prime/R(P).\eqno(15)$$
By the theorem that the mean square is greater than or equal to the
square of the mean modules $I\geq 1$. 

Now in a self-similar distribution of field $I$ would be independent
of $Z$ so we shall suppose that at heights well above the base at
$z=0$, $I$ depends on $Z$ only weakly if at all. Using (13) for
${\Big\langle}B^2_z{\Big\rangle}$ in (10) we find
$$\pi \left [R(P)\right ]^2 = P\left [I/(2\pi p_0)\right ]^{1/2},\eqno(16)$$
so neglecting the variation of $\log I$,
$$d\ln R(P)/d {\rm ln}P = 1/2.\eqno (17)$$
Now let ${\Big\langle}B^2_\phi{\Big\rangle}{\Big /}\bar{B}^2_\phi = J.$
Then from (10) and (12)
$$\int^P_0 \frac{\Phi(P^\prime)}{Z(P^\prime)} dP^\prime = 8\pi \left
(\frac{\pi p_0}{J}\right)^{1/2}R(P).$$
Differentiating with respect to $P$ and using (17) and (16),
$$\Phi(P)/Z(P) = 4\pi(\pi p_0/J)^{1/2}R(P)/P = (8I/J)^{1/2}/R(P),$$
where we have neglected the variation of $J$ with height.
Hence $$\frac{Z(P)}{R(P)} = \frac{\Phi(P)}{(8I/J)^{1/2}},$$
or, since $\Phi(P) = \Omega (P)t$,
$$Z(P) = (8I/J)^{1/2}\Omega(P)R(P)t,\eqno(18)$$
so the maximum height of each field line grows linearly in time at a
velocity related to the disk's circular velocity, and the area
occupied by the field at height $Z(P)$ is proportional to $P$ as given
by (16).

To proceed further we need to determine $\Omega(P)$, which is one of
the inputs of our model since it is determined by the distribution of
the poloidal flux over the disk as well as the disk's rotation.

We shall take our accretion disk to be Keplerian outside some radius
$R_0$ so that the circular velocity
$$V=V_0(R_0/R_1)^{1/2} \hspace{1cm} (R_1 > R_0)$$
however, such a law cannot continue down to $R_1=0$ as it would lead
to infinite velocities.  We shall therefore take circular velocities
of the form $$V=V_0{\Big [ }1+\left( R_1/R_0\right) ^2{\Big ]}^{-1/4}.$$
My original accretion disk model of quasars (Lynden-Bell 1969) relied
on shearing and reconnecting magnetic fields giving stresses to
provide the torques which allowed the material to lose its angular
momentum and be accreted.  It gave magnetic fields $B\propto R^{-5/4}$,
and more recent accretion disk models (e.g., Shakura \& Sunyaev
1973, 1976;  Tout \& Pringle 1992), which rely on the less reliable
$\alpha-$viscosity, give the same $R$ dependence for magnetic fields.
However, again the field must reach some finite value as $R_1$ becomes
small so we shall take on the disk $z=0$
$$B_z = B_0 (1+(R_1/R_0)^2)^{-5/8}.\eqno (19)$$
This gives a convenient flux function
$$P=F_0 \frac{8}{3}\left\{\left [1+\left (R_1/R_0\right)^2\right
]^{3/8}-1\right\} \hspace{1cm}{\rm for}\ P\geq F_0\eqno(20)$$
where $F_0 = \pi R^2_0 B_0$.  We shall assume there is a total
poloidal flux $F$ so $P \leq F$.  Eliminating $R_1/R_0$ we obtain 
$$V=V_0\left [1+(3/8) P/F_0\right ]^{-2/3}\hspace{1cm}({\rm for}\ P\leq
F)$$
or writing $\Omega_0 = V_0/R_0$ we have 
$$V/R_1 = \Omega = \Omega_0 \left [1+(3/8) P/F_0\right ]^{-2/3} 
\left\{\left [1+\frac{3P}{8F_0}\right ]^{8/3}-1\right\}^{-1/2},\eqno(21)$$ 
where the last factor is $R_0/R_1$ rewritten and we have neglected
$\Omega(R_2)$ compared with $\Omega(R_1)$.  

For small $P$,  $\Omega = \Omega_0 (P/F_0)^{-1/2}.$

We   are now in a position to determine the shape of the magnetic
cavity occupied in the magnetic field.  From (16) $P\propto \left [R(P)\right ]^2$ so we get $\Omega$ in terms of $R(P)$ from (18).  For
$R(P)\leq b$
$$\Omega = \Omega_0\left [1+(3/8)\left (R(P)/a\right )^2\right
]^{-2/3}\left\{\left [ 1 + \frac{3}{8}\left
(\frac{R(P)}{a}\right)^2\right ]^{8/3}-1\right \}^{-1/2},
\eqno(22)$$ where $a^2 = \pi^{-1}F_0\left [I/(2\pi p_0)\right ]^{1/2},$
which is constant and $b^2 = (F/F_0)a^2$.  Thus from (18)
writing  $x = R(P)/a$ we have for $ x\leq b/a$:

$$Z(P) = (8I/J)^{1/2} a\Omega_0 t \left [1+(3/8) x^2\right
]^{-2/3}\left \{ \left [1+(3/8)x^2\right]^{8/3}-1 \right \}^{-1/2}x.
 \eqno(23)$$ 
The factors after the first bracket $\rightarrow 1$ for
$x$ small.  $Z(P)$ gives the height and $R(P)$ alias $x$ gives the
radius of the cavity occupied by field so a plot of this equation gives
the shape at each time.  $R(P)$ has a maximum at $b$ and at lower $Z$
it remains at $b$ down to $Z \sim a$ where the field configuration
becomes that of Figure 3.  The shape of the cavity is plotted as
Figure 4.  
\begin{figure}[t]
\plotone{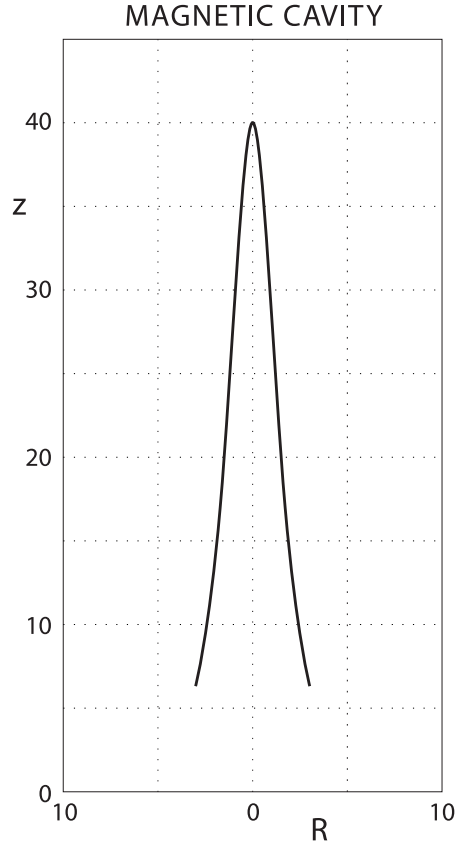}
\caption{The magnetic cavity occupied by the field becomes
systematically taller with time but its lateral extent remains the same.}
\end{figure}

The stability of the towers is an interesting subject that cannot be
treated here.  However, some general guidance can be given.  Long thin
towers in compression are unstable to sideways buckling like an Euler
strut.  If by contrast they are in tension, they are stable to
such buckling.  However, even a liquid is stable under compression if
it is surrounded by other liquid at the same pressure.  Thus if we
imagine a long thin tower of liquid it will only buckle if asked to
support a pressure in excess of the ambient pressure of the
surrounding liquid.  From this example it is reasonable to expect our
magnetic towers to buckle only if they experience a net compressive
force greater than the ambient pressure.  With
$(\langle B^2_\phi\rangle -
\langle B^2_z\rangle )/8\pi - p_0 =0$ our towers are just stable by
that criterion.  However, we have neglected two other effects.
Firstly, magnetic field is buoyant, so if $p_0$ decreases with height
this should require a net tension from the magnetic stresses.
Secondly, ram pressure at the head of any advancing jet will require
some extra compression along the jet.  As these two effects act in
opposite directions we must leave the stability of the towers
undecided.  Finally, it is not clear that all real jets are stable to
buckling; that of 3C273 appears bent on milliarcsecond scales.

\section{Conclusions}
In reality jets are dynamic whereas the towers we have calculated are
static.  In so far as the ram pressure and other inertial effects can
be neglected, evolution through growing towers gives a realizable
model.  But such models which are very useful for guidance and
understanding, should not be considered as giving the exact shapes to
be expected from fully dynamic jets.  However, they do supply an
interesting and provocative answer to the question ``Why are there
jets at all?''  Theories of jet models that fail to answer that question but
assume an imposed flux of material from their base may be
missing this point.  Our picture also gives a good explanation of why
jets advance at speeds directly related to the circular or escape
velocity of the inner parts of the disks from which they emerge.

\end{document}